\documentclass[amsmath,showpacs,superscriptaddress,amssymb,twocolumn]{revtex4-1}

	\usepackage[english]{babel}
	\usepackage{graphicx}
	\usepackage{float}
	\usepackage{dcolumn}
	\usepackage{bm}
	\usepackage{subfigure} 
	\usepackage{float}
	
	\usepackage{epsf}

\begin{document}
		\title{Chimera states in Star Networks}
		
		\author{Chandrakala Meena \footnote{email : chandrakala@iisermohali.ac.in}}
			\affiliation{Indian Institute of Science Education and Research (IISER) Mohali, Knowledge City, SAS Nagar, Sector $81$, Manauli PO $140$ $306$, Punjab, India}
		\author{K. Murali{\footnote{email : kmurali@annauniv.edu}}}
			\affiliation{Department of Physics, Anna University, Chennai $600$ $025$,India}
		\author{Sudeshna Sinha{\footnote{email : sudeshna@iisermohali.ac.in}}}
			\affiliation{Indian Institute of Science Education and Research (IISER) Mohali, Knowledge City, SAS Nagar, Sector $81$, Manauli PO $140$ $306$, Punjab, India}
	
		{\footnotetext{ \protect\vspace*{2.5cm}} }

	\begin{abstract}
	
	  We consider star networks of chaotic oscillators, with all end-nodes connected only to the central hub node, under diffusive coupling, conjugate coupling and mean-field type coupling. We observe the existence of chimeras in the end-nodes, which are identical in terms of the coupling environment and dynamical equations. Namely, the symmetry of the end-nodes is broken and co-existing groups with different synchronization features and attractor geometries emerge. Surprisingly, such chimera states are very wide-spread in this network topology, and large parameter regimes of moderate coupling strengths evolve to chimera states from generic random initial conditions. Further, we verify the robustness of these chimera states in analog circuit experiments. Thus it is evident that star networks provide a promising class of coupled systems, in natural or human-engineered contexts, where chimeras are prevalent.
	
	\end{abstract}
	
	\maketitle
	\section{Introduction}
	
	Chimera states have been extensively studied over the last decade in natural and artificial networks of coupled identical complex systems, in fields ranging from physics and chemistry to biology and engineering \cite{chimera,chimera15,chimera14,chimera1,chimera2,chimera4,chimera5,chimera6,chimera7,chimera8,chimera9,chimera10,chimera11,chimera12,chimera13,chimera16,chimera17,chimera18,chimera19,chimera20}. At the outset, Kuramoto and his colleagues \cite{chimera,chimera14,chimera15} first noticed that in a system of non-locally coupled identical phase oscillators, in a ring configuration, the system spontaneously broke the underlying symmetry and split into synchronized and desynchronized oscillator groups. Namely, there emerged a state where coherent and incoherent sets of oscillators co-existed. This state was dubbed a chimera state \cite{chimera1}, as it was reminiscent of the greek mythological creature composed of incongruous parts. Subsequently this fascinating phenomena has been observed in a variety of non-locally coupled systems, such as time delayed systems \cite{chimera12}, Josephson junction arrays \cite{chimera16}, electrochemical systems \cite{chimera17} and uni-hemispheric sleep in certain animals \cite{chimera18}.
	In recent years chimera states have also been observed experimentally in optical analogs of coupled map lattices \cite{chimera9}, Belousov-Zhabotinsky chemical oscillator systems \cite{chimera8}, two populations of mechanical metronomes \cite{chimera19} and modified time delay electronic circuit systems \cite{chimera20}.  
	
	Till now chimera states have been reported primarily in networks that have a regular ring topology, where oscillators are coupled in a non-local \cite{nonlocal,chimera12,chimera4} or global fashion \cite{chimera5,chimeraglobal}. In this work we will show how chimera states also emerge in {\em oscillator networks with a star topology}. The star configuration is one where the network has a central hub position and all other nodes are linked to this node \cite{chimera13}. This configuration arises extensively in computer networks, where every node connects to a central computer, and the central computer act as a server and the peripheral devices act as clients. Further, a star-like structure is a primary motif in scale-free networks, which have been reported to arise in wide-ranging phenomena \cite{scalefree}. 
	
	Here we will show the extensive existence of chimeras in the end-nodes of the star network, which are identical in terms of the coupling environment and dynamical equations. We will demonstrate how the symmetry of the end-nodes is broken and co-existing groups with different dynamical behaviour emerge. Interestingly we find that such chimera states are very wide-spread in this network topology, and large parameter regimes of coupling strengths typically yield a chimera state. We also confirm the existence of robust chimera states in analog circuit experiments.

	\section{Star Networks of Chaotic Oscillators}

	Here we study the dynamics of a star network of $N$ identical nonlinear oscillator systems. In such networks there is one central hub node (labelled by site index $i = 1$) and $N-1$ environmentally identical peripheral end-nodes connected to the central node (labelled by node index $i=2, \dots N$). One can also interpret this system as a set of uncoupled oscillators connected to a common drive. The focus of this study is the dynamical patterns arising in the $N-1$ identical end-nodes of this network. 
	In order to establish the generality of our results, we consider three different coupling forms: (a) diffusive coupling (b) conjugate coupling and (c) mean-field coupling. We give below the general dynamical equations for the different coupling forms.	
	First, we consider standard diffusive coupling through similar variables, given by:
	\begin{eqnarray}
	\dot{x_{i}}&=&f_x (x_i, y_i, z_i) + \sum_{j=1}^N K_{ij}(x_{j}-x_{i}) \\ \nonumber
	\dot{y_{i}}&=&f_y (x_i, y_i, z_i) \\ \nonumber
	\dot{z_{i}}&=&f_z (x_i, y_i, z_i)
	\end{eqnarray}
	Here coupling matrix element for central node $i=1$ is $K_{1j}=k/2$ when $j \neq 1$, and for the end-nodes $i=2, \dots N$, $K_{i1}=k/2$ and zero otherwise. The coupling strength is given by $k$.
	Then we consider the {\em conjugate coupling} \cite{chimera3} given as:
	\begin{eqnarray}
	 \dot{x_{i}}&=&f_x (x_{i},y_{i},z_{i})+\sum_{j=1}^N K_{ij} (y_{j} - x_{i} )\\ \nonumber 
     \dot{y_{i}}&=&f_y (x_{i},y_{i},z_{i})\\ \nonumber
	 \dot{z_{i}}&=&f_z (x_{i},y_{i},z_{i})
	\end{eqnarray}
	Lastly, we also consider a mean-field type of coupling, where the dynamics of the central node is given by:
	\begin{eqnarray}
	\dot{x_1} &=& f_x (x_1, y_1, z_1) + \frac{k}{2} (x_m -  x_1) \\ \nonumber							
	\dot{y_1} &=& f_y (x_1, y_1, z_1) \\ \nonumber						
	\dot{z_1} &=& f_z (x_1, y_1, z_1)
	\label{expt_meanfield}
	\end{eqnarray}
	where $x_m = \frac{1}{N-1} \sum_{j=2, \dots N} x_j$ is the mean field of the end-nodes. The dynamics of the end-nodes $i = 2, \dots N$ is given by:
	\begin{eqnarray}
	\dot{x_i}&=&f_x (x_i, y_i, z_i) + \frac{k}{2} (x_1 - x_i) \\ \nonumber
	\dot{y_i}&=&f_y (x_i, y_i, z_i) \\ \nonumber	
	\dot{z_i}&=&f_z (x_i, y_i, z_i)
	\label{expt_coup2}
	\end{eqnarray}

	For the local dynamics at the nodes, we take two prototypical chaotic systems that have widespread relevance in modelling phenomena ranging from lasers to circuits. First we consider the R{\"ossler} type oscillator at node $i$, given by the form:
	\begin{eqnarray}
	f_x (x_{i}, y_i, z_i)&=&-[\omega_{i} + \epsilon({x_{i}}^{2}+{y_{i}}^{2})]{y_{i}} - {z_i}\\ \nonumber
	f_y (x_{i}, y_i, z_i)&=&[\omega_{i} + \epsilon({x_{i}}^{2}+{y_{i}}^{2})]{x_{i}} + a y_{i}\\ \nonumber
	f_z (x_{i}, y_i, z_i)&=&b + z_{i}(x_{i} - c)
	\end{eqnarray} 
	in Eqns. 1-4. For each node, $\omega_{i}+\epsilon({x_{i}}^{2}+{y_{i}}^{2})$ is close to the angular velocity of the $i^{th}$ oscillator, perturbed by amplitude ${x_{i}}^{2}+{y_{i}}^{2}$ when $\epsilon \neq 0$. Here we take the parameter values to be: $a=0.15$, $b=0.4$, $c=8.5$, $\omega_{1}=\omega_{2}=\omega_{3}=0.41$ and $\epsilon=0.0026$ \cite{chimera3}, yielding a chaotic attractor. 
	We also consider the Lorenz system at the nodes given by:
	\begin{eqnarray}
	f_x (x_{i}, y_i, z_i)&=&\sigma( y_{i} - x_{i}) \\ \nonumber
	f_y (x_{i}, y_i, z_i)&=&(r-z_{i}) x_{i} - y_{i} \\ \nonumber
	f_z (x_{i}, y_i, z_i)&=&x_{i}y_{i} - \beta z_{i}
	\end{eqnarray}
	in Eqns. 1-4. With no loss of generality we consider the parameters of the local system to be $\sigma=10$, $r=28$ and $\beta=8/3$, yielding double-scroll attractors.
	We study both these chaotic systems, coupled in star network configuration, through the different coupling forms given above. A wide range of coupling strengths, in networks of size ranging from $3$ to $100$ oscillators is investigated. The principal observations of the patterns arising in these networks, from generic random initial states, are described below.
	
	\subsection{Dynamical Patterns for Coupled R{\"o}ssler Oscillators}
		\begin{figure}[H]
			\centering
                \includegraphics[width=0.8\linewidth]{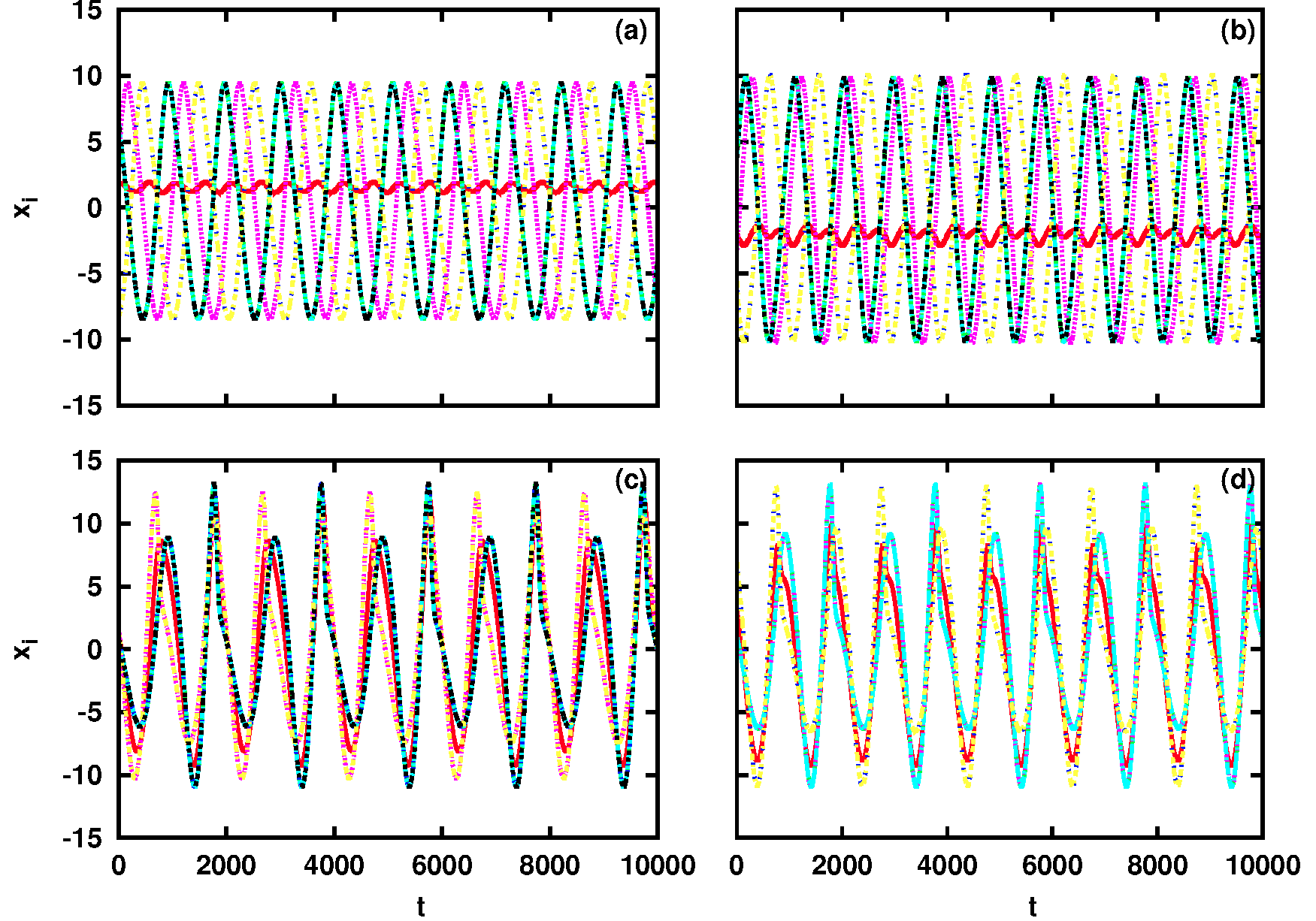}
                \caption{Time evolution of the $x$ variable for oscillators in distinct synchronized groups in a coupled R{\"o}ssler systems in a star network of (a) $10$ conjugately coupled nodes (cf. Eqn. 2) for coupling strength $k=0.22$; (b) $100$ conjugately coupled nodes (cf. Eqn. 2) for coupling strength $k=0.18$; (c) $10$ diffusively coupled nodes (cf. Eqn. 1) for coupling strength $k=0.14$; and in (d) $100$ diffusively coupled nodes (cf. Eqn. 1) for coupling strength $k=0.14$. In all these figures the identical end-nodes, that split into $3$ groups in (a) and (b), and into $2$ groups in (c) and (d) are marked with dotted and dashed lines of different colors. The one distinct central hub node is also shown (in solid red) for comparison.}
				\label{time_rossler}
		\end{figure}
	We find that as coupling strength increases, the end-nodes go from a de-synchronized state to a completely synchronized state, via a large coupling parameter regime yielding chimera states. In the representative examples of chimera states displayed in Fig. \ref{time_rossler}, the $9$ identical end-nodes of the network of $10$ conjugately coupled oscillators (Fig \ref{time_rossler}a), clearly split into $3$ clusters, with two synchronized clusters having $4$ oscillators each, and $1$ oscillator being distinct from both these synchronized groups. For the case of $100$ conjugately coupled nodes in Fig. \ref{time_rossler}b, the $99$ identical end-nodes again cluster into $2$ synchronized groups of size $49$ each, and $1$ oscillator is uncorrelated to either group. Notice that the central node settles down to low amplitude oscillations, while the end-nodes exhibit large amplitude oscillations, with each group having a different phase with respect to another. For the coupling strengths presented in the figure, regular low-period oscillations emerge in the end-nodes, though the constituent oscillators were chaotic. 
	
	For the case of $10$ diffusively coupled R{\"o}ssler oscillators, the identical end-nodes split into $2$ synchronized clusters of sizes $5$ and $4$  (Fig \ref{time_rossler}c), while the end-nodes of a star network of $100$ diffusively coupled oscillators split into $2$ synchronized clusters of size $53$ and $46$ oscillators  (Fig \ref{time_rossler}d). Here the central node and the end-nodes all exhibit large amplitude oscillations of higher periodicity.

\begin{figure}[H]
		\centering              
		\includegraphics[height=0.99\linewidth,width=0.95\linewidth,angle=0]{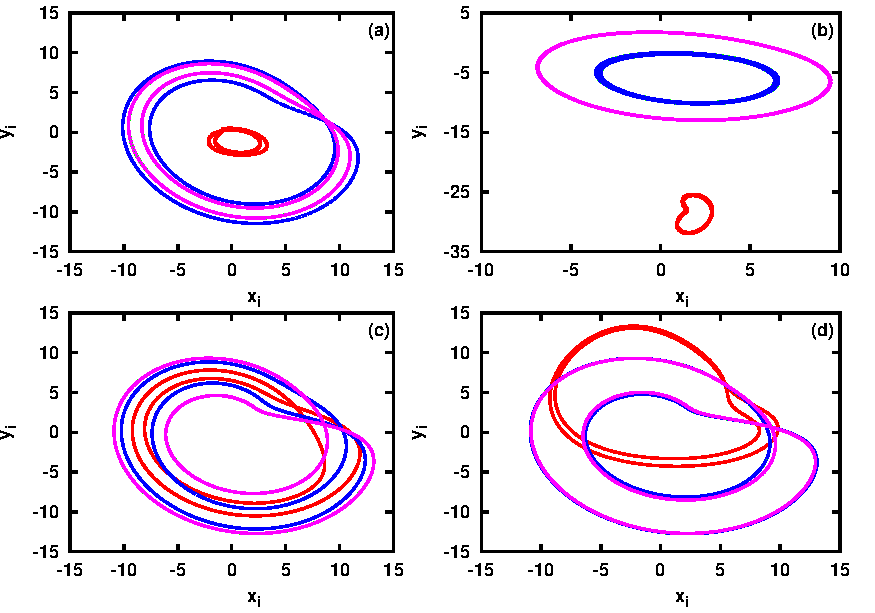}
        \caption{Phase portraits of conjugately coupled R{\"o}ssler systems in a star network with (a) $10$ nodes and coupling strength $k=0.12$, (b) $100$ nodes and coupling strength $k=0.22$. Phase portraits of diffusively coupled R{\"o}ssler systems in a star network with (c) $10$ nodes and coupling strength $k=0.14$, (d) $100$ nodes and coupling strength $k=0.14$.  In all these figures the distinct end-node clusters are marked in blue and cyan. The central hub node is also shown (in solid red) for comparison.}
		\label{phase_rossler}
	\end{figure}

\begin{figure}[H]
		\centering
		\includegraphics[width=0.75\linewidth,angle=0]{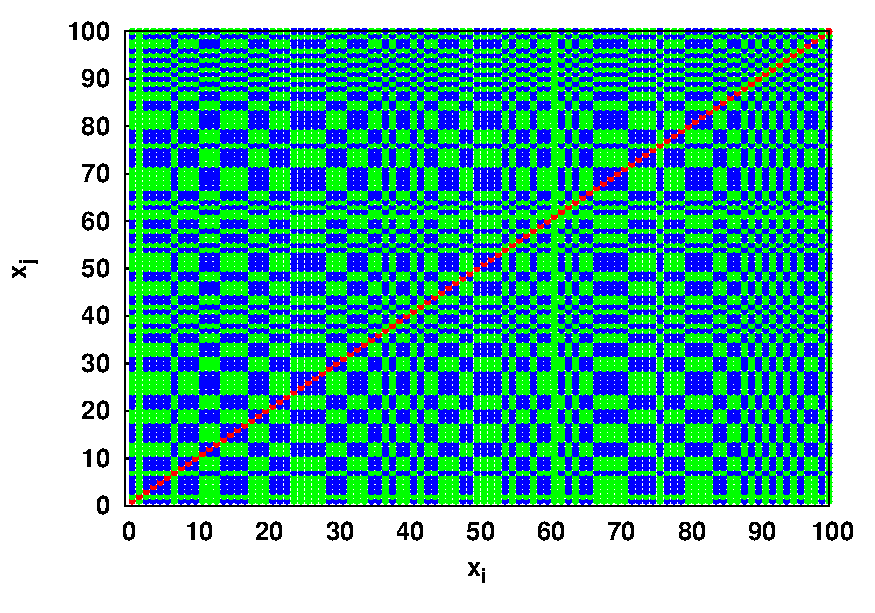}
		\caption{A matrix displaying the state of synchronization of nodes $i$ and $j$ in a star network of conjugately coupled R{\"o}ssler systems ($i,j =2, \dots N$). The blue color indicates that the nodes are synchronized and the green that they are desynchronized. Here coupling strength $k=0.24$ and system size $N=100$. The presence of a synchronized group of nodes, along-side a desynchronized set, can be clearly seen.}
		\label{syncmatrix}
	\end{figure}

	Fig. \ref{phase_rossler} shows the oscillatory patterns of the end-nodes for the distinct synchronized groups that emerge from generic random initial states. For instance, it is evident from Fig. \ref{phase_rossler} (a-d) that sub-sets of the end-nodes display very different attractor geometries, though they have identical dynamical equations. So from Figs. \ref{time_rossler} and \ref{phase_rossler} it is clearly evident that chimera states emerge in the end-nodes of the star network. Further, Fig. \ref{syncmatrix} shows the state of synchronization of the different end-nodes $i=2, \dots N$ at some representative instant of time. demonstrating the co-existence of synchronized and de-synchronized groups among the identical $N-1$ peripheral nodes in the star network. Note that there is no space ordering of the node index $i$ of the end-nodes. So the (de)synchronized nodes in a cluster are not ``contiguous'', as is usual in regular lattice topologies.  

	\subsection{Dynamical Patterns for Coupled Lorenz systems}
\begin{figure}[H]
		\centering
		\includegraphics[width=0.7\linewidth]{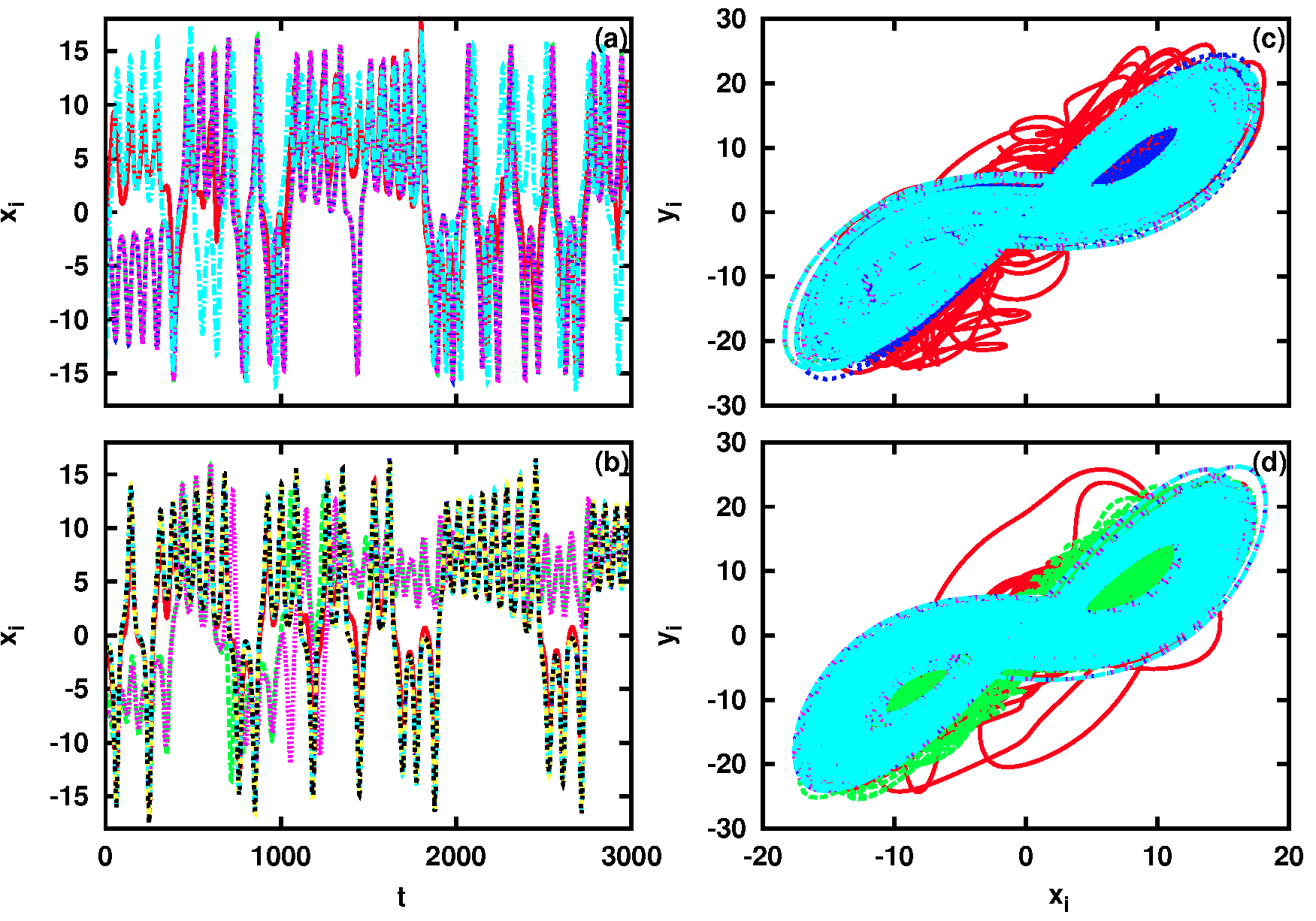}
         \caption{Temporal patterns (a-b) and their corresponding phase portraits (c-d) for a star network of coupled Lorenz system with (a) $10$ conjugately coupled nodes and coupling strength $k=2.9$, yielding $2$ synchronized clusters of size $6$ and $3$; (b) $10$ diffusively coupled nodes for coupling strength $k=13$, yielding one synchronized cluster of size $7$, and a desynchronized cluster of size $2$.}
		\label{lorenz}
	\end{figure}

\begin{figure}[H]
		\centering
		\includegraphics[width=0.9\linewidth,angle=0]{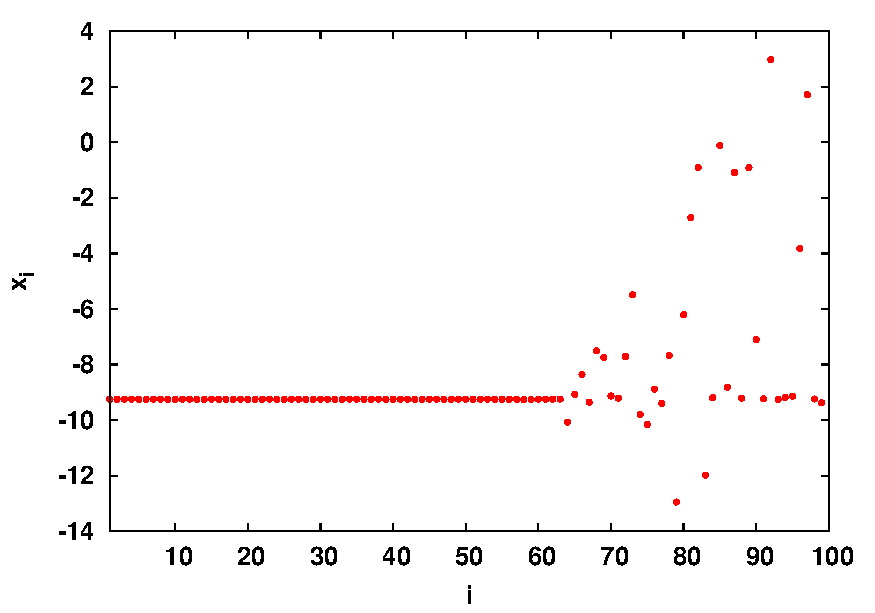}
		\caption{State $x_{i}$ of the end-nodes of a star network of conjugately coupled Lorenz systems ($i=2, \dots N$), at an instant of time. Here coupling strength $k=1.98$ and system size $N=100$. The presence of a synchronized group of nodes, along with a desynchronized set, can be clearly seen.}
		\label{state}
	\end{figure}

	Here again we find that as coupling strength increases, the end-nodes go from a de-synchronized state to a completely synchronized state, via a large coupling parameter regime yielding chimera states. We display some representative patterns from the chimera states in Fig. \ref{lorenz}. It is clearly evident from these that the identical end-nodes split into different dynamical groups, thereby breaking symmetry. Some of these groups consists of synchronized nodes and some are clusters of de-synchronized elements, as seen from Fig. \ref{state}. Further, it is also evident from Fig. \ref{lorenz} that in addition to different synchronization properties, the groups also yield different attractor geometries.

\begin{figure}[H]
		\centering
		\includegraphics[height=0.7\linewidth,width=0.95\linewidth,angle=0]
		{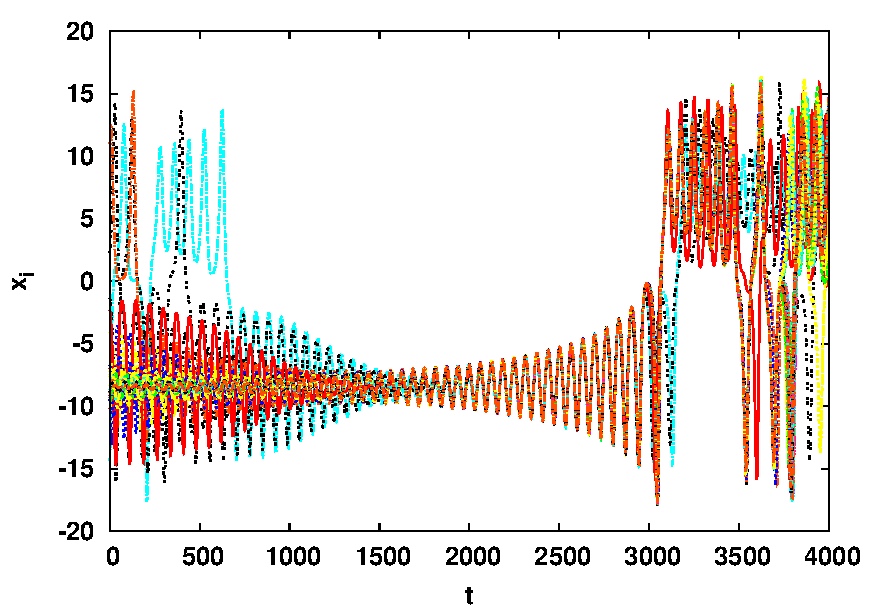}
		\caption{Temporal patterns of the end-nodes of a star network of diffusively coupled Lorenz systems displaying a breathing chimera state. Here coupling strength $k=5.0$ and system size $N = 100$.}
		\label{breathing1}
	\end{figure}

\begin{figure}[H]
		\centering
		\includegraphics[height=0.6\linewidth,width=0.75\linewidth,angle=0]
		{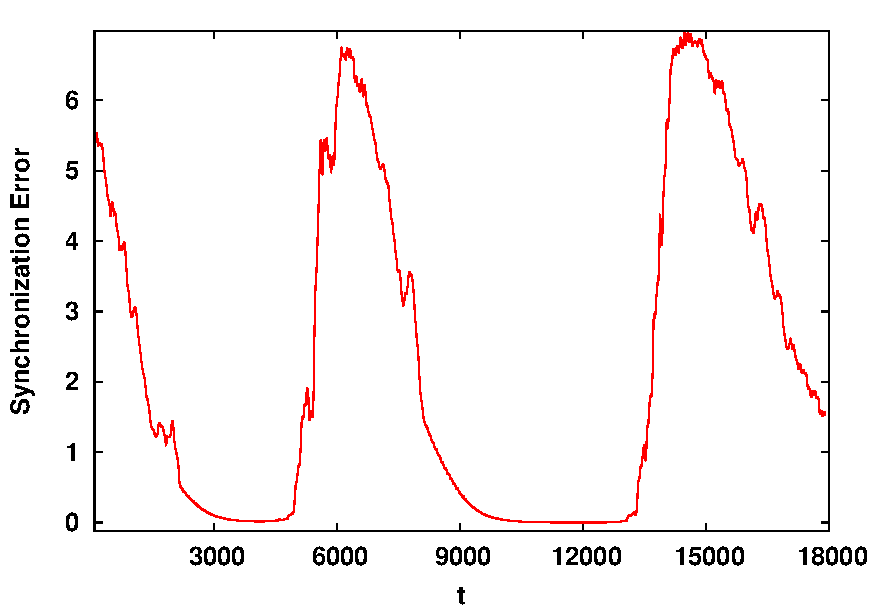}
		\caption{Synchronization error of the end-nodes of a star network of diffusively coupled Lorenz systems as a function of time (namely the standard deviation of $x_i$, $i=2, \dots N$, at an instant of time). Here coupling strength $k=5.12$ and system size $N=100$. It is clearly evident from the oscillating synchronization error that the end-nodes move in and out of synchronization.}\label{breathing2}
	\end{figure}

	Further we find that the incoherent state may be desynchronized at the same level (stable chimera) or yield an oscillating incoherent group which goes in and out of synchronization, namely a {\em breathing chimera} \cite{chimera22}. Such a breathing chimera state is displayed in Figs. \ref{breathing1}-\ref{breathing2}.
	The occurence of breathing chimera states is more common in the coupled Lorenz system than in coupled R{\"o}ssler systems. In fact breathing chimeras were also observed in Lorenz systems coupled in a ring configuration in earlier studies \cite{chimera6}.	
\subsection{Prevalence of chimera states }
	In order to quantify the probability of obtaining chimera states from random initial states we calculate the fraction of initial conditions leading to co-existing synchronized and desynchronized states in the end-nodes, in a large sample of random initial states. This provides an estimate of the basin of attraction of the chimera state, and indicates the prevalence of chimeras in this system. So this measure is important, as it {\em allows us to gauge the chance of observing chimeras without fixing special initial states}.

\begin{figure}[H]
		\centering		\includegraphics[height=0.99\linewidth,width=0.95\linewidth]{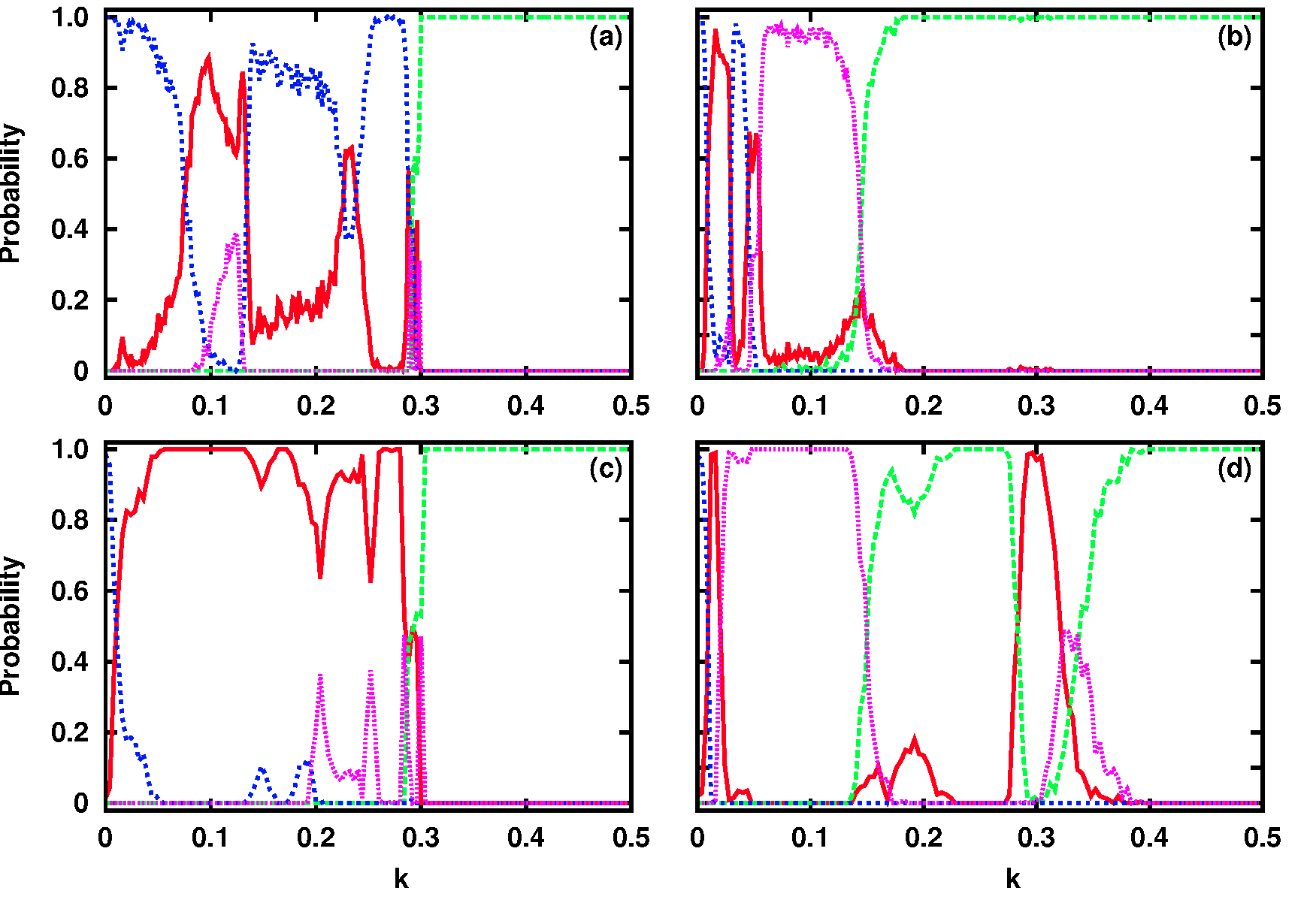}
		\caption{Probability of obtaining chimera states (red), synchronized clusters (magenta), fully synchronized states (green), and completely de-synchronized states (blue) in star networks of coupled R{\"o}ssler systems, for the following cases: $10$ nodes under (a) conjugate coupling and (b) diffusive coupling; $100$ nodes under (c) conjugate coupling and (d) diffusive coupling.}\label{prob1}
	\end{figure}

\begin{figure}[H]
		\centering		\includegraphics[height=0.8\linewidth,width=0.8\linewidth]{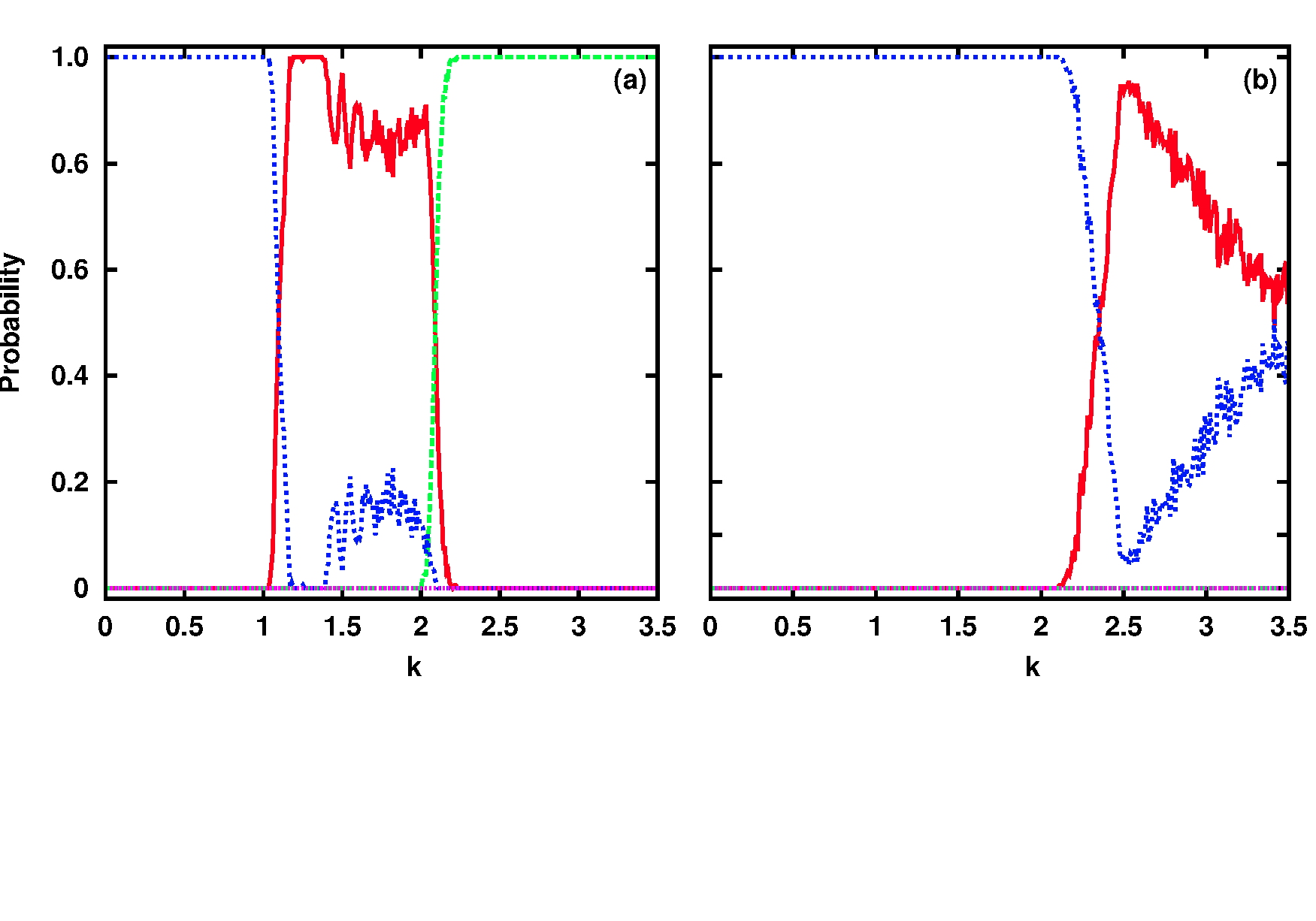}
		\caption{Probability of obtaining chimera states (red), synchronized clusters (magenta), fully synchronized states (green), and completely de-synchronized states (blue) in star networks of coupled Lorenz systems, of $100$ nodes, under (a) conjugate coupling and (b) regular diffusive coupling.}\label{prob2}
	\end{figure}

	Figs. \ref{prob1} and \ref{prob2} display this quantity for star networks of R{\"o}ssler and Lorenz systems. It is clearly evident from these figures that there exists extensive regimes of coupling parameter space where the probability of obtaining a chimera state is close to one. This quantitively establishes the prevalence of chimeras in the end-nodes of nonlinear oscillators coupled in star configurations. Also notice that larger networks yield larger basins of attraction for the chimera state. Further, the figures show that {\em conjugate coupling yields larger parameter bands with high prevalence of chimera states}.
	\section{Experimental Verification of Chimera States}
\begin{figure}[H]
	\centering
	\includegraphics[width=0.4\linewidth]{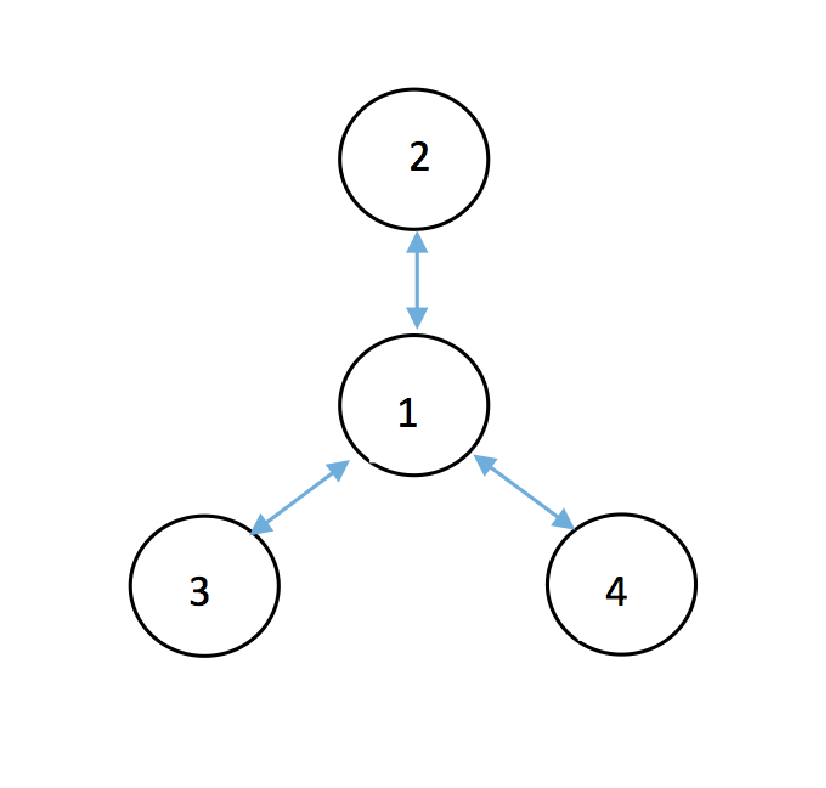}
	\caption{Schematic of the star network realized through analog simulation circuits. Here block $1$ represents the central node and blocks $2$,$3$, and $4$ represent end-nodes.}
	\label{schematic}
	\end{figure}

\begin{figure}[!ht]
	  \centering
	  \includegraphics[width=0.8\linewidth]{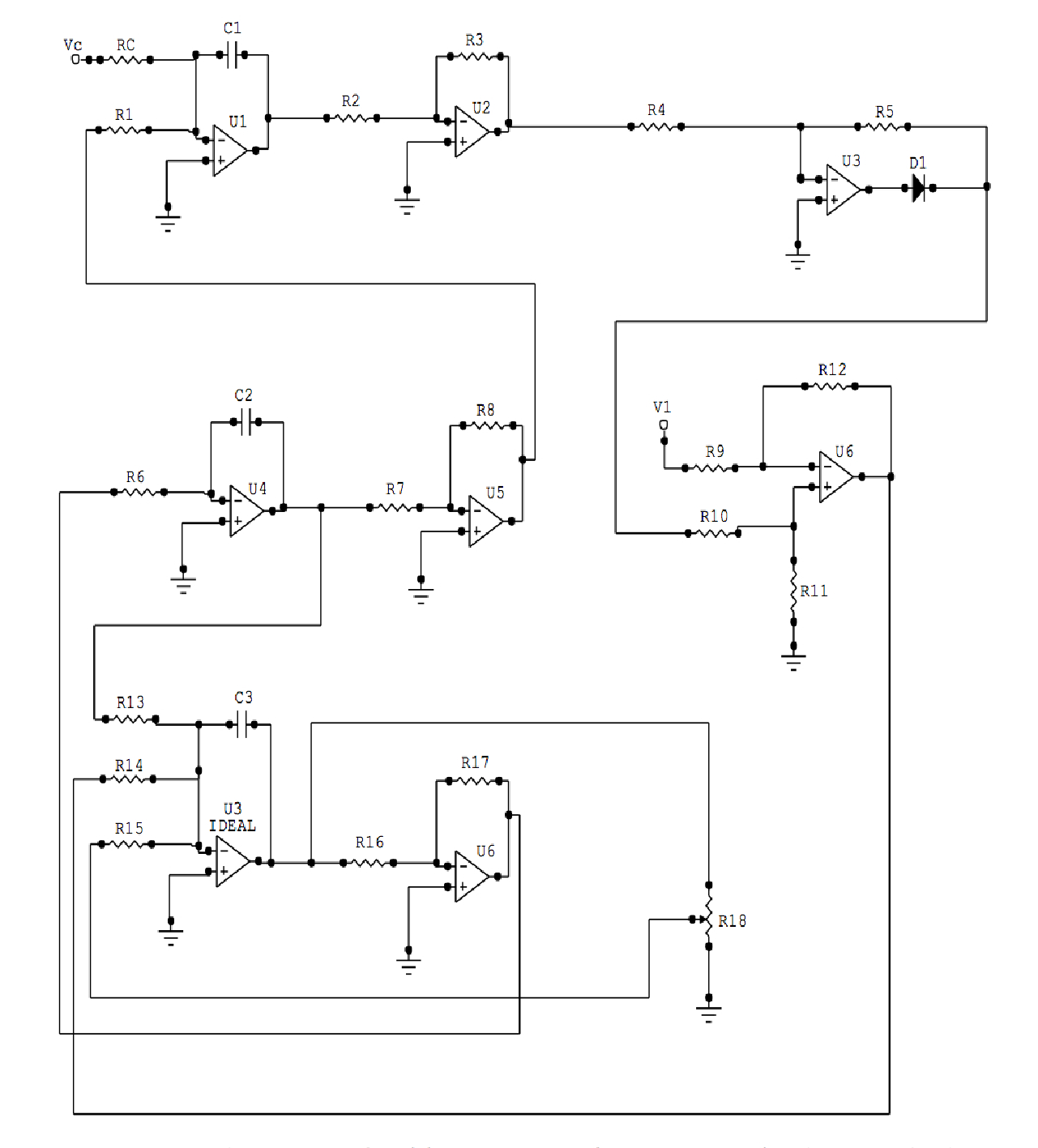}
	  \caption{Circuit implementation of Eqns. 8-9 using op-amps ($AD712$ or $\mu A741$).  The external voltage supply ($V1$) is $+1.0 V$.  The capacitors are $10nF$. Resistors $R4$ and $R5$ = $1 k \Omega$, the variable resistor is $10 k \Omega$, and the rest are $100 k \Omega$.  The diode is $IN4148$. $Vc$ is the input coupling signal.}
	  \label{circ2}
	\end{figure}

\begin{figure}[!1ht]
	\centering
	\includegraphics[width=0.8\linewidth]{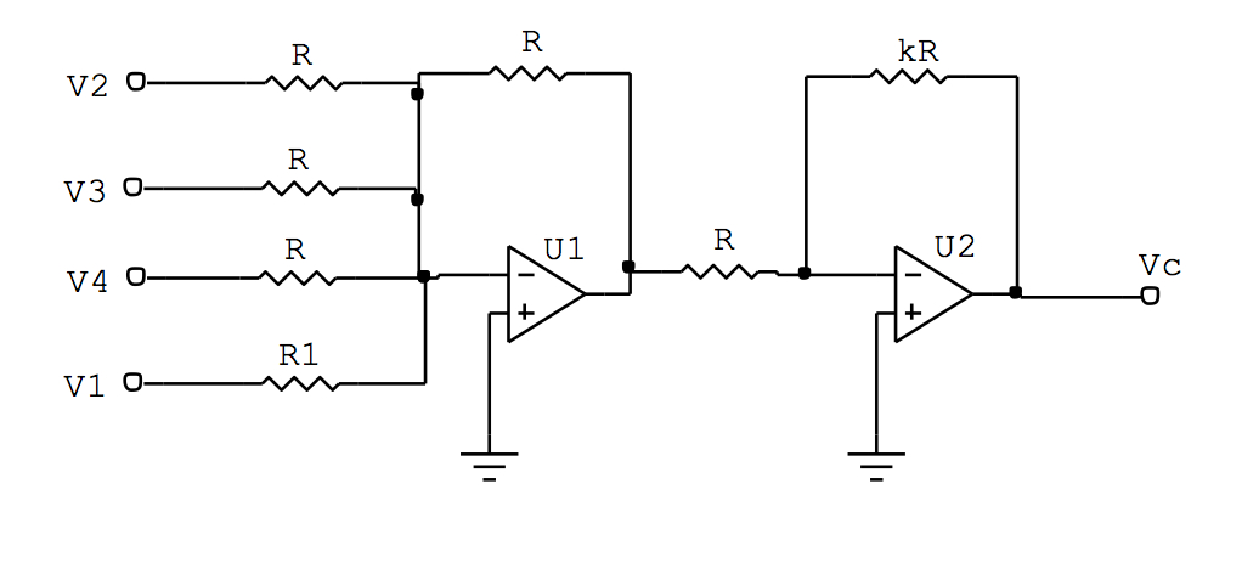}
	\caption{Circuit used to implement coupling between the end-nodes and the central node. All op-amps are $AD712$ or $\mu A741$. Resistor $R1 = 100 k \Omega$, and the rest are $300 k \Omega$.  $Vc$ is the output coupling voltage, and $V1 = -x_1$ is signal generated by the central node. Signals $V2$, $V3$ and $V4$ corresponding to $x_2$, $x_3$ and $x_4$, are generated by using three more circuit copies of Fig.11.}
	\label{circ3}
	\end{figure}

\begin{figure}[!ht]
	\centering
	\includegraphics[width=0.8\linewidth]{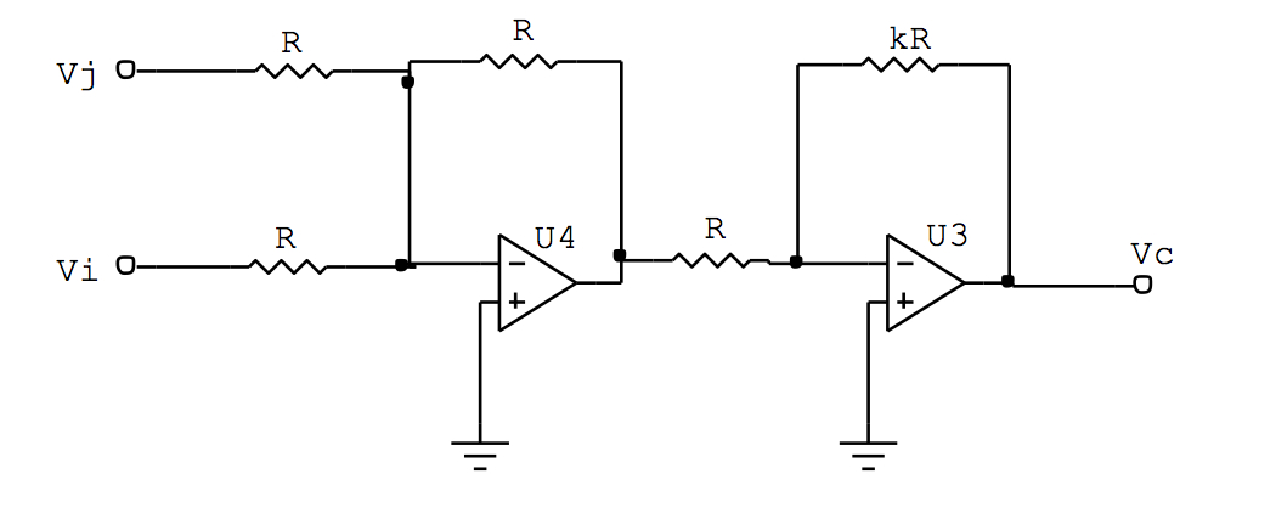}
	\caption{Circuit used to implement coupling between the central node and the end-nodes.  All op-amps are $AD712$ or $\mu A741$ and the resistors are $100 k \Omega$. $Vc$ is the output coupling signal, $Vj = x_1$ from the central node and $Vi = -x_i$ from the end-nodes.}\label{circ4}
	\end{figure}

\begin{figure}[htbp]
	    \centering
		\includegraphics[height=0.45\linewidth]{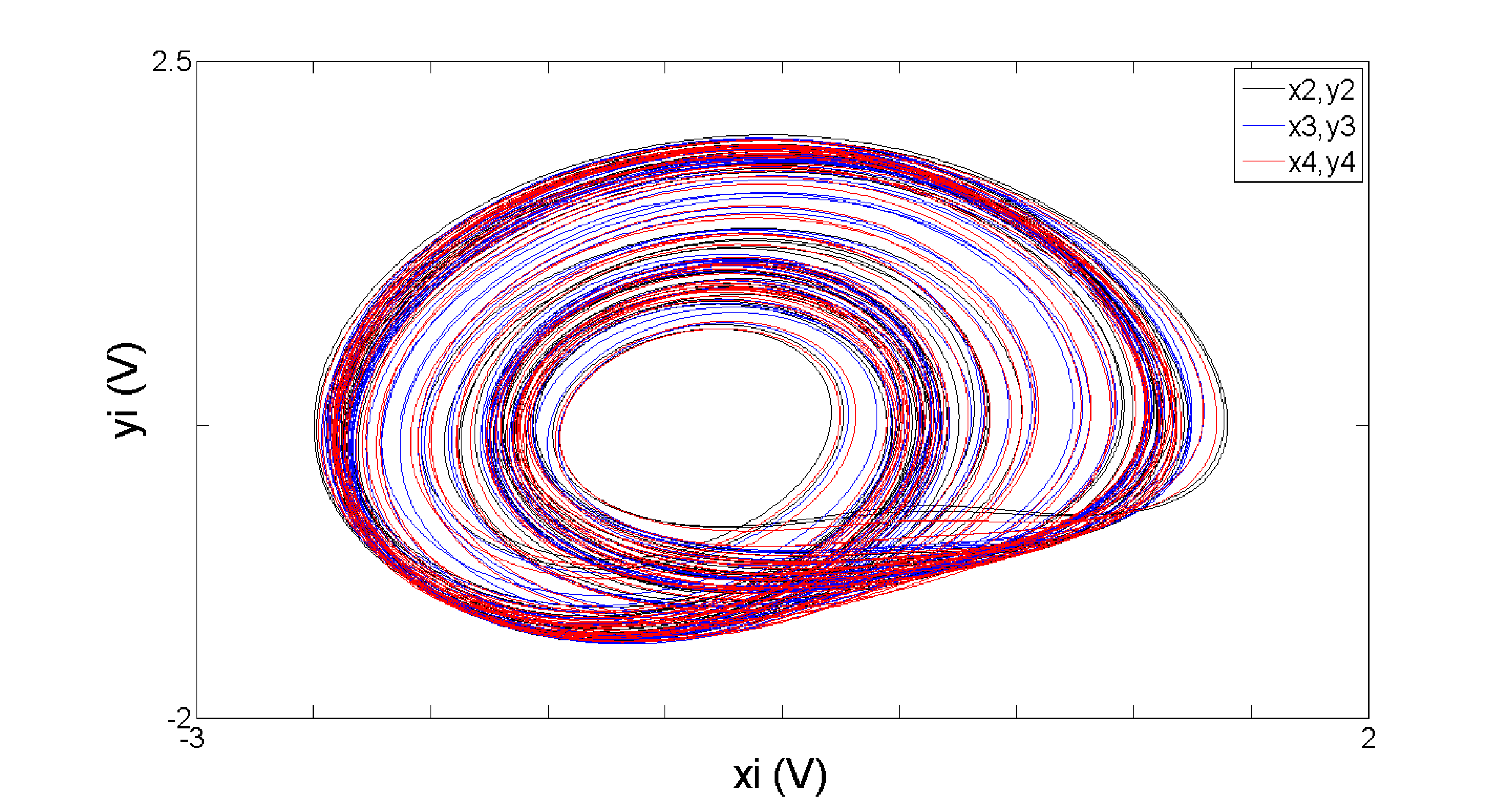}(a)
		\includegraphics[height=0.45\linewidth]{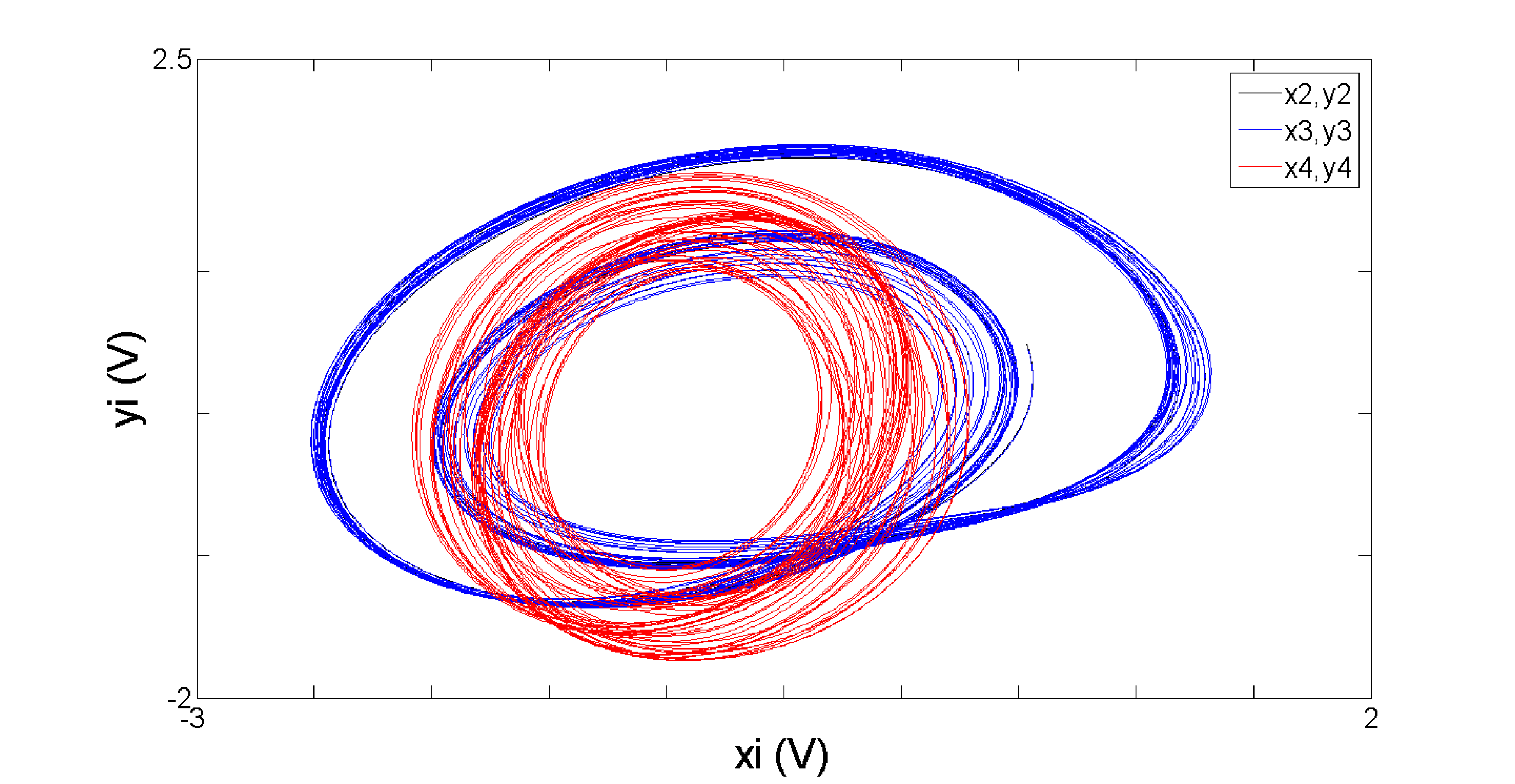}(b)
		\includegraphics[height=0.45\linewidth]{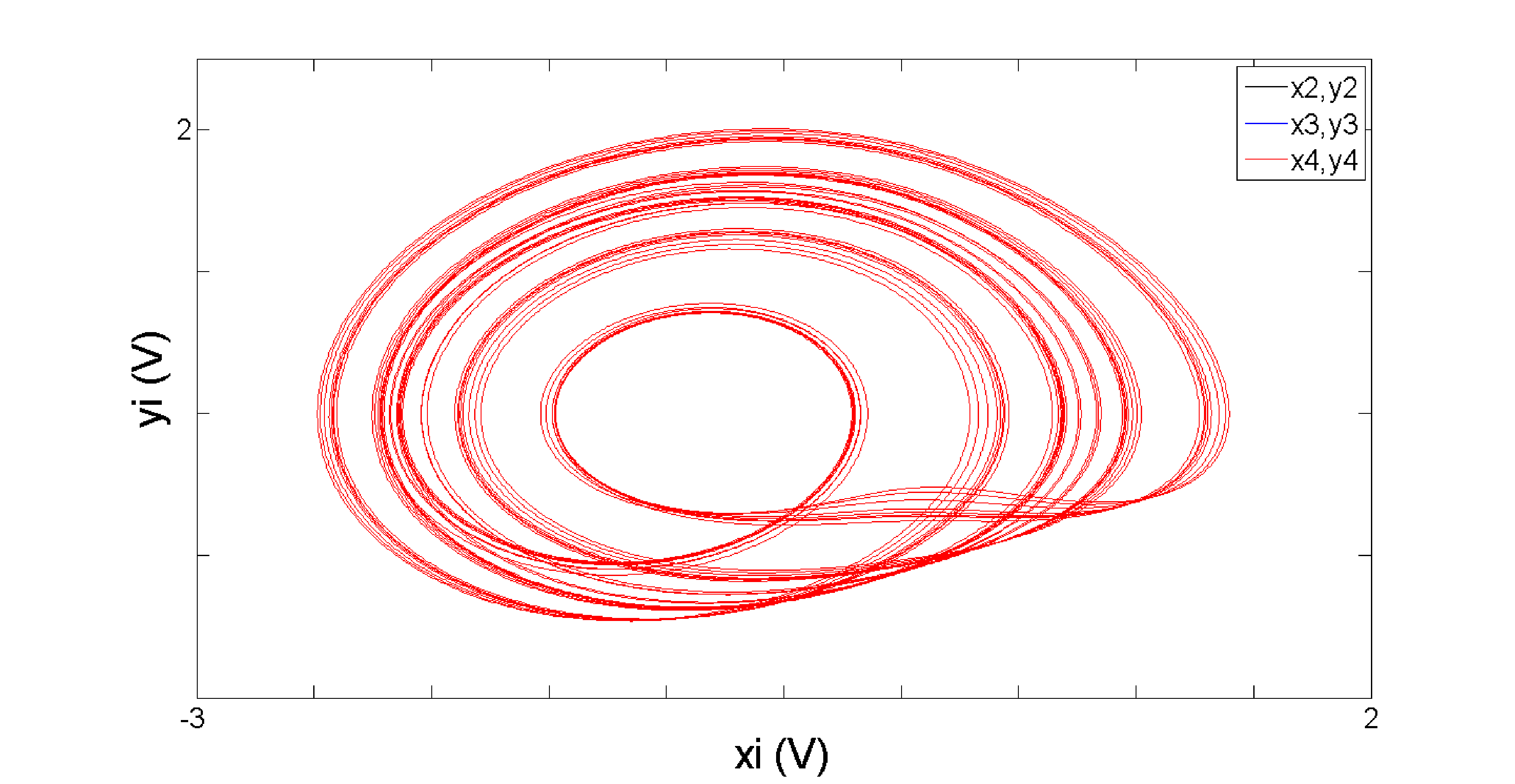}(c)
		\caption{Phase portrait of attractors in the $x_i - y_i$ plane, generated from chaotic R{\"o}ssler-type oscillator circuits represented by Eqn.~\ref{expt}, that are diffusively coupled in a star network, for coupling strengths: (a) $k = 0.1$ yielding an unsynchronized state, (b) $k = 1.0$ yielding a chimera-like state and (c) $k = 2.0$ yielding a synchronized state. In all these figures the oscillations of the end-nodes in the star network are marked in different colours. The central hub node oscillator dynamics is not shown.}
		\label{expt1}
	\end{figure}

\begin{figure}[H]
		\centering
		\includegraphics[width=1.15\linewidth,height=1.3\linewidth]{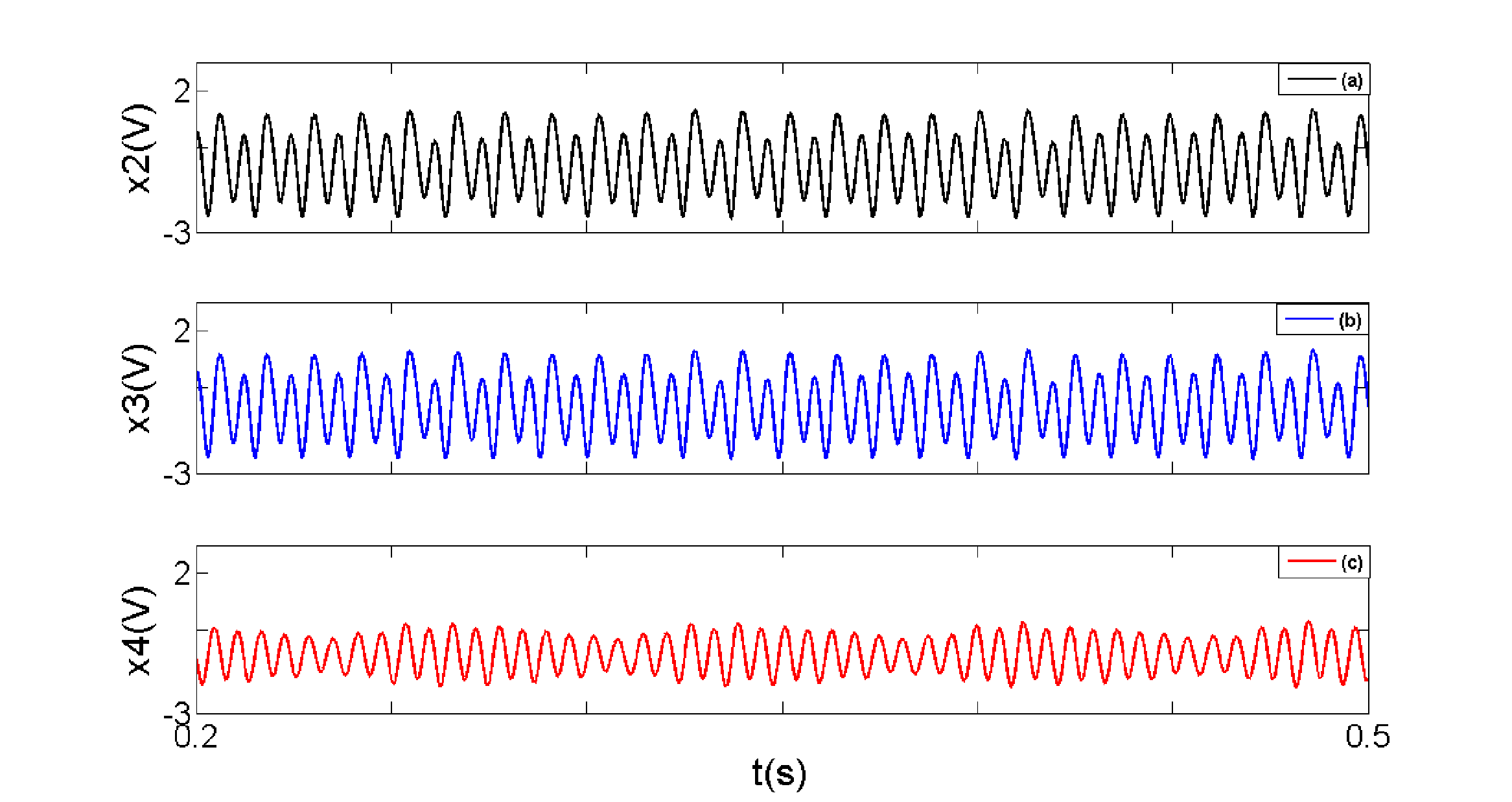}
		\caption{Times series of the $3$ end-nodes (a-c) of $4$ diffusively coupled chaotic R{\"o}ssler-type circuits, represented by Eqn~\ref{expt} with coupling strength $k = 1.0$. Clearly $2$ nodes (a,b) are synchronized, while one (c) is distinct from this group.}
		\label{expt2}
	\end{figure}

\begin{figure}[H]
		\centering
		\includegraphics[height=0.45\linewidth]{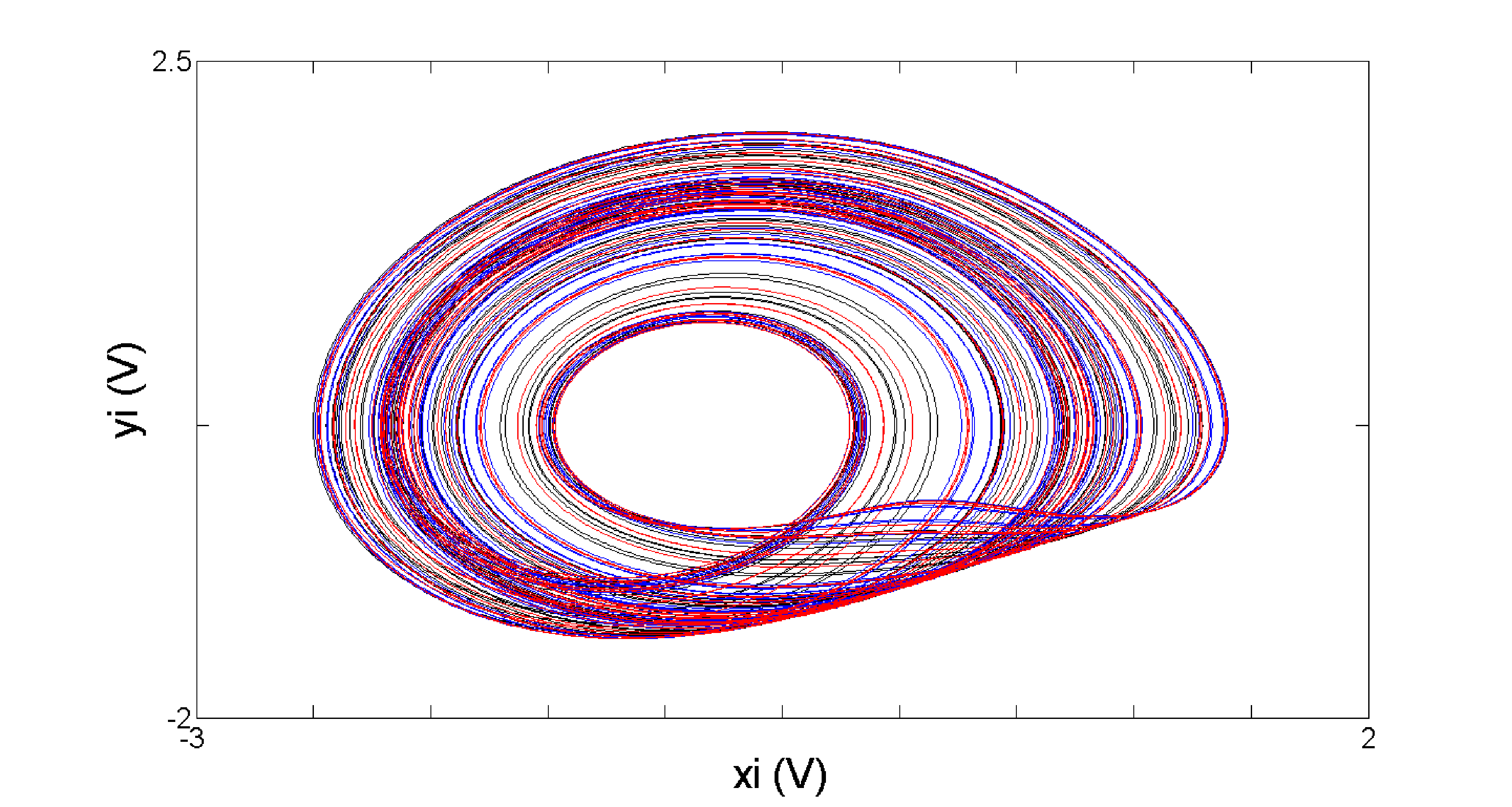}(a)
		\includegraphics[height=0.45\linewidth]{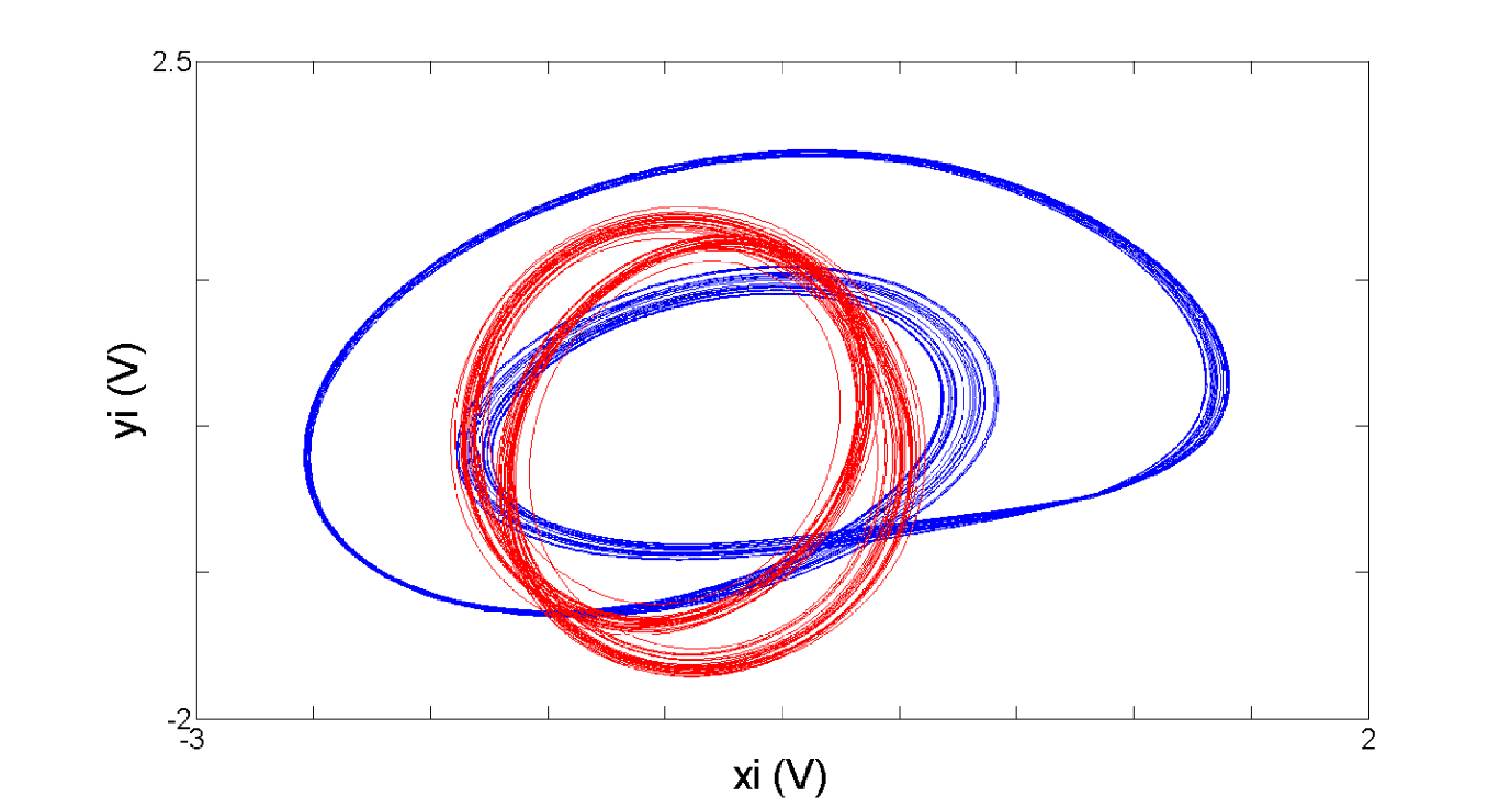}(b)
		\includegraphics[height=0.45\linewidth]{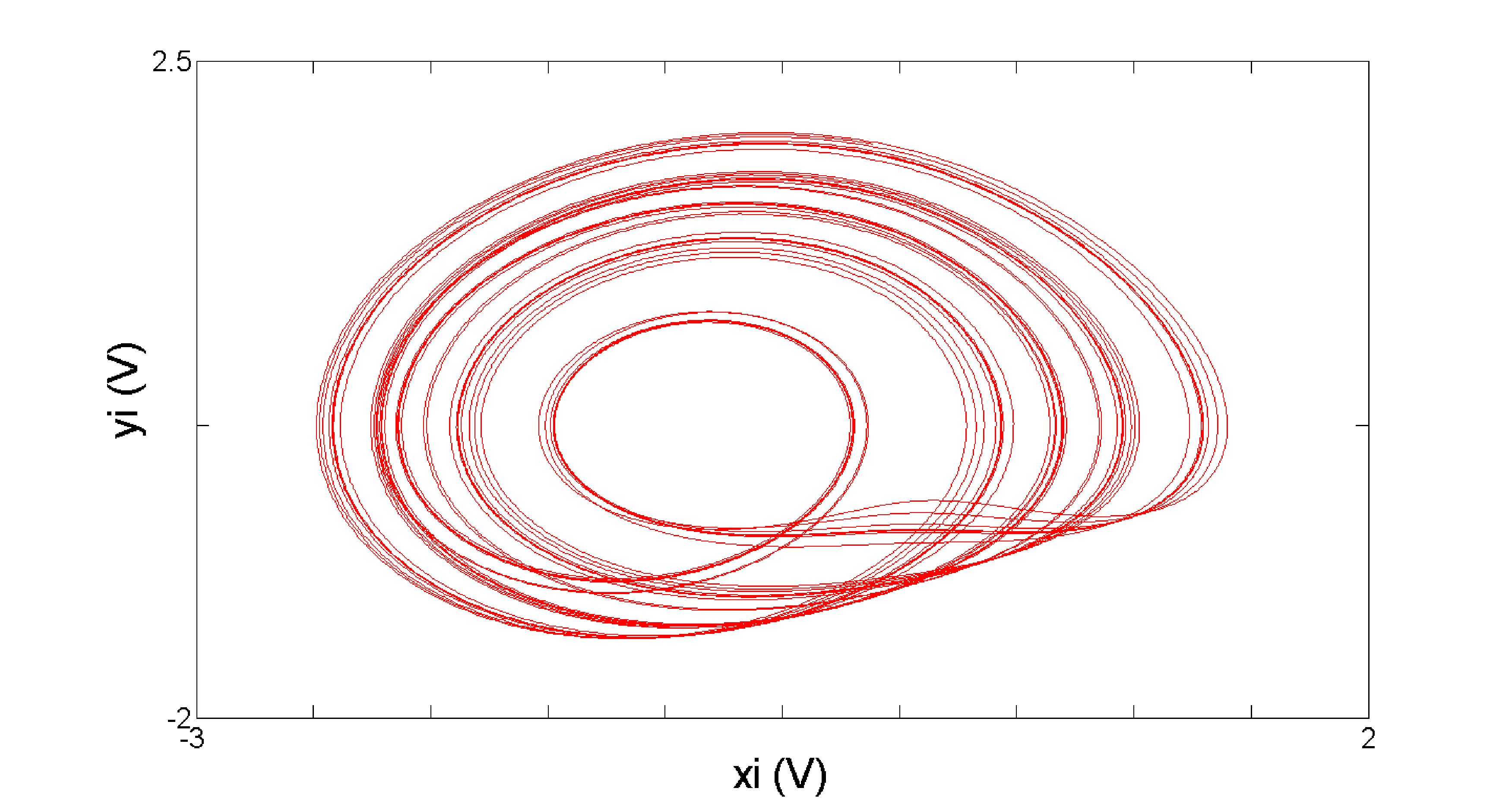}(c)
		\caption{Phase portrait of attractors in the $x_i - y_i$ plane, generated from chaotic R{\"o}ssler-type oscillator circuits represented by Eqn.~\ref{expt}, coupled via mean-field  in a star network (cf. Eqns. 3-4), for coupling strengths: (a) $k = 0.1$ yielding an unsynchronized state, (b) $k = 1.0$ yielding a chimera-like state and (c) $k = 2.0$ yielding a synchronized state. In all these figures the oscillations of the identical end-nodes in the star network are marked in different colours. The central hub node oscillator dynamics is not shown.}\label{expt3}
		\end{figure}

\begin{figure}[H]
		\centering		
		\includegraphics[height=0.51\linewidth,width=0.65\linewidth]{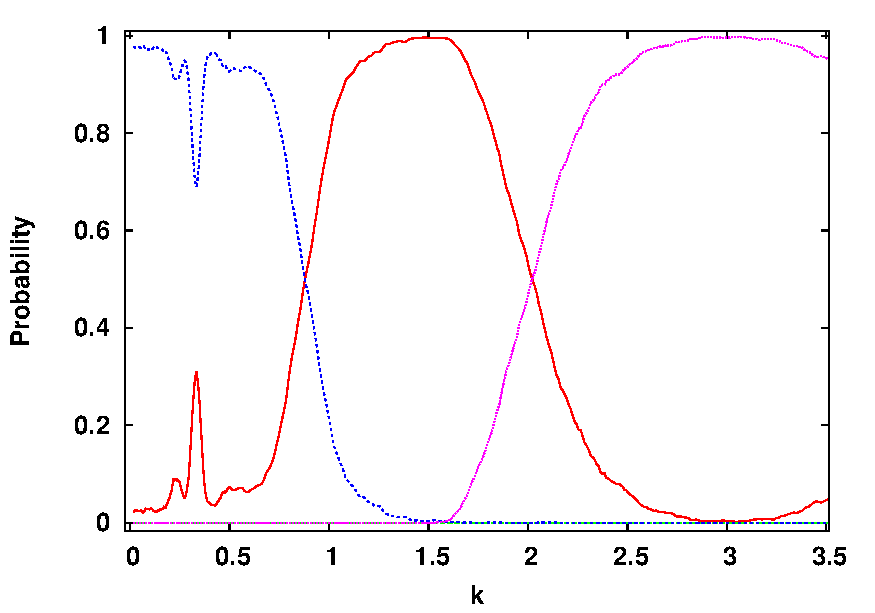}
		\caption{Probability of obtaining chimera states (red), synchronized clusters (magenta), fully synchronized states (green), and completely de-synchronized states (blue) in star networks of coupled R{\"o}ssler systems with mean-field type coupling  (cf. Eqns. 3-4) for a network of $100$ nodes.}\label{prob_meanfield}
	\end{figure}

		Now we establish the robustness of these chimera states in experimental situations by demonstrating the occurrence of chimera states in star networks of coupled nonlinear oscillators, evolving from {\em generic initial states}. Specifically, we consider a circuit implementation of a chaotic R{\"o}ssler-type oscillator at the nodes, represented by the equation \cite{expt}:
	\begin{equation}
	\frac{d^3x}{dt^3} = -A \frac{d^2x}{dt^2} - \frac{dx}{dt} \pm (|x| - 1)  	
	\label{expt}				
	\end{equation}
	This equation illustrates a jerk type chaotic system \cite{expt}, and an analog simulation circuit of this equation can be carried out with standard operational amplifiers and diodes. The details of a straightforward circuit implementation of Eqn.~\ref{expt} can be found in Ref. \cite{expt}. We then go on to set up $4$ diffusively coupled oscillators, with parameter $A=0.58$ in Eqn. \ref{expt} such that the oscillators individually exhibit chaotic dynamics. 
	Specifically, we experimentally study the star network given schematically in Fig. \ref{schematic}, where the central node evolves as: 
	\begin{eqnarray}
	\dot{x_1}&=&y_1 + \frac{k}{2}\left( x_2 + x_3 + x_4 - 3x_1 \right)\\ \nonumber
	\dot{y_1}&=&z_1\\ \nonumber						
	\dot{z_1}&=&-A z_1 - y_1 \pm (|x_1| - 1)
	\label{hub}
	\end{eqnarray}
	The evolution of the identical end-nodes is given by:
	\begin{eqnarray}
	\dot{x_i}&=&y_i + \frac{k}{2} (x_1 - x_i) \\ \nonumber
	\dot{y_i}&=&z_i\\ \nonumber						
	\dot{z_i}&=&-A z_i - y_i \pm (|x_i| - 1)
	\label{endnodes}
	\end{eqnarray}
	where $i = 2,3,4$ and $k$ is the coupling co-efficient. 
	Fig. \ref{circ2} depicts the electronic analog circuit to implement the central node or end-nodes of Eqns. \ref{hub}-\ref{endnodes}. If the circuit of Fig. \ref{circ2} acts as a central node, then the output voltages from op-amp $U1$ and $U2$ correspond to $-x_1$ and  $x_1$ of Eqns. 8-9. If we use the circuit of Fig. \ref{circ2} for the end-nodes, then they generate the signals $V2 = x_2$, $V3 = x_3$ and $V4 = x_4$ at output of $U2$.  For the central node circuit, the input voltage $Vc$ is the coupling voltage signal generated from the circuit of Fig. \ref{circ3}. In this case, the $Vc$ signal corresponds to $(k/2)(x_2+x_3+x_4 - 3x_1)$. Fig. \ref{circ4} depicts the circuit used to implement the coupling between central node to the end-nodes. In this case, $Vc$ signal corresponds to  $(k/2)(x_1  - x_i)$. 
		Representative circuit simulation results are displayed in Fig. \ref{expt1}, where phase-portraits in the $x_i - y_i$ plane are displayed for different coupling strengths $k$. One clearly notices that for low coupling strength (e.g. $k = 0.1$ in Fig.\ref{expt1}a) the end-nodes show completely unsynchronized oscillations. For large coupling strength (e.g. $k = 2.0$ in Fig.\ref{expt1}c), as anticipated, the end-nodes exhibit complete synchronization. However, for moderate coupling strengths (e.g. $k = 1.0$ in Fig.\ref{expt1}b) the $3$ identical end-nodes split into two groups, where two of them are synchronized and one is not, thus exhibiting a chimera-like state. The time series of this state is shown in Fig. \ref{expt2} to further illustrate the broken symmetry of the three identical end-nodes in the star network. Note that we have no control over the initial state in the experiment, and these states evolve from generic random initial conditions.
		Further, in order to check the generality of the results, we also investigate the mean-field type of coupling given by Eqns. 3-4. Fig. \ref{expt3} displays representative phase-portraits in the $x_i - y_i$ plane for different coupling strengths $k$. Again one finds that for low coupling strengths (e.g. $k = 0.02$ in Fig.\ref{expt3}a) the end-nodes are completely unsynchronized, while for high coupling strengths (e.g. $k = 2.0$ in Fig.\ref{expt3}c) they are completely synchronized. However, in a large window of moderate coupling strengths (e.g. $k = 1.0$ in Fig.\ref{expt3}b) the $3$ identical end-nodes split into two groups, where two of them are synchronized and one is not, thus exhibiting a chimera-like state. Also note the different geometries of the dynamical state in the two groups.
		Lastly, we estimate the probability of obtaining the chimera state in the star network with mean-field coupling by finding, through numerical simulations, the fraction of initial states that evolve to chimera states. The results are displayed in Fig. \ref{prob_meanfield}, and it is clear that this form of coupling yields a large parameter regime where the typical initial state gives rise to a chimera state in the end-nodes.
	\section{Conclusions}
		In summary, we have investigated star networks of diffusively and conjugately coupled nonlinear oscillators, with all end-nodes connected only to the central hub node. Though the end-nodes are identical in terms of the coupling environment and dynamical equations, they yielded chimera states. Namely, the symmetry of the end-nodes was broken and co-existing groups with different synchronization features and attractor geometries emerged. 
        We estimated the basin of attraction of chimera states by evaluating the fraction of initial states that evolve to a chimera state, in a large sample of random initial conditions. This measure showed that in extensive regimes of coupling parameter space the probability of obtaining a chimera state is close to one. Further, we established the robustness of these chimera states in analog circuit experiments. The experimental verifications incorporated both diffusive coupling and mean-field type coupling for the central node. 
        Thus it is clearly evident from our numerical and experimental investigations that large parameter regimes of moderate coupling strengths yield chimera states from generic random initial conditions in this network topology. So star networks provide a promising class of coupled systems, in natural or human-engineered contexts, where chimeras are pervasive.
        
        {\bf Acknowledgements}\\
			CM would like to acknowledge the financial support from DST INSPIRE Fellowship, India. Further CM acknowledges stimulating discussions and help in programming from Pranay Deep Rungta.

	\end{document}